\documentclass[epj]{svjour}

\usepackage{graphicx}
\usepackage{fancyhdr}
\def\makeheadbox{{
 }}
\def\mytitle{Search for TeV in Flavour } 
\def\myauthors{George W.S. Hou}    
\def\mytype{Review}
\def\mysession{\myauthors}

\begin{document}
\title{Search for TeV Scale Physics in Heavy Flavour Decays}
\author{George W.S. Hou
}                     
%
%
\institute{Department of Physics, National Taiwan
 University, Taipei, Taiwan 10617, R.O.C.
}
%
\date{}
\abstract{
 The subject of heavy flavour decays as probes for physics
 beyond the TeV scale is covered from the experimental perspective.
 Emphasis is placed on the more traditional Beyond the Standard Model
 topics that have potential for impact in the short term,
 with the physics explained.
 We do unabashedly promote our own phemonenology work.
\PACS{
      {13.20.He}{Decays of bottom mesons}   \and
      {12.60.-i}{Models beyond the standard model}
     } 
} 
\maketitle

As humans we aspire to reach up to the heavens, to reach beyond
the veiling clouds of the v.e.v. scale. The conventional high
energy approach, such as the LHC, is like Jack climbing the bean
stalk up into the clouds, where impressions are that the Higgs
boson is just floating in a lower cloud close by, but then maybe
not. However, ``Jack" may not have to actually climb the bean
stalk: quantum physics allows him to stay on Earth, and let
virtual ``loops" do the work. It is in this way that flavour
physics offers probes of the TeV scale, at reduced costs.

To illustrate the potential impact, let us entertain a
hypothetical ``What if?" question, by forwarding to the recent
past. On July 31, 2000, the BaBar experiment announced at Osaka
conference the low value of $\sin2\beta \sim 0.12$. The Belle
value for the equivalent $\sin2\phi_1$ was slightly higher, but
also consistent with zero. Within the same day, a theory paper
appeared on the arXiv~\cite{KN00}, entertaining the implications
of the low $\sin2\beta$ value. A year later, however, both BaBar
and Belle claimed the observation of $\sin2\beta/\phi_1 \sim 1$,
which turn out to be consistent with SM expectations. But, {\it
what if it stayed close to zero?} Well, you would have heard more
about it: a definite large deviation from SM! Even beforehand, one
expected from indirect data that in SM context,
$\sin2\beta/\phi_1$ had to be nonzero.

With $\beta/\phi_1 = -\arg V_{td}$ in SM, it is instructive to
recall that $B^0$--$\bar B^0$ mixing was discovered by the ARGUS
experiment 20 years ago, which was the first clear indication that
$m_t$ is heavy. This illustrates the power of flavour loops as
probes into high scales. The nondecoupling of the top quark from
the box diagram, $M_{12}^d \propto m_t^2 (V_{tb}V_{td}^*)^2$,
illustrates the {\it Higgs affinity} of heavy SM quarks, i.e.
$\lambda_t \sim 1$. At the same time, it is this {\it Higgs
affinity that allows us to probe the CPV $\beta/\phi_1$ phase at
the B factories.}

With $b\to d$ transitions seemingly consistent with SM, i.e. no
discrepancy in the CKM triangle
\begin{eqnarray}
V_{ud}^*V_{ub} + V_{cd}^*V_{cb} + V_{td}^*V_{tb} = 0,
 \label{eq:btodTri}
\end{eqnarray}
what about $b\to s$ transitions? This will be our starting point
and the main theme.

\section{\boldmath CPV in $b \to s$ with Boxes and Penguins}
\label{btos}

With $\tau\to \mu$ echoes, CPV in $b \leftrightarrow s$
transitions is the current frontier. We focus on four topics.


\subsection{\boldmath $\Delta{\cal S}$}
 \label{DeltaS}

\begin{figure}[b!]
\hskip0.6cm
\includegraphics[width=0.40\textwidth,height=0.28\textwidth,angle=0]{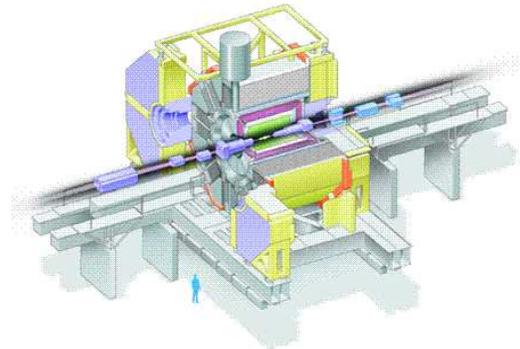}
\caption{Schematic picture of Belle detector.}
 \label{fig:Belle}       
\end{figure}

The B factories were built to measure time-dependent CP violation
(TCPV) in $B^0\to J/\psi K_S$ mode. Besides reconstructing the
final state which is a CP eigenstate, one needs to {\it tag} the
other $B$ meson flavour ($B^0$ or $\bar B^0$), and measure both
the $B$ decay vertices. The BaBar and Belle
 (illustrated schematically in Fig.~\ref{fig:Belle}
detectors are rather similar, differing basically only in the
particle identification detector (PID) used for flavour tagging.
Both the PEP-II and KEKB accelerators started in 1999. By 2001,
KEKB/Belle outran PEP-II/BaBar in luminosity.

With TCPV in $B^0\to J/\psi K_S$ measured by 2001, attention
quickly turned to the $b\to s$ penguin modes, where a virtual
gluon is emitted from the virtual top quark in the vertex loop.
Take $B^0 \to \phi K_S$ for example, where the virtual gluon pops
out an $s\bar s$ pair. The $b\to s$ penguin loop amplitude is
practically real within SM, just like the tree level $B^0\to
J/\psi K_S$. This is because $V_{us}^*V_{ub}$ is very suppressed.
Thus, SM predicts
\begin{eqnarray}
{\cal S}_{\phi K_S} \cong \sin2\phi_1/\beta,
 \label{eq:SphiK}
\end{eqnarray}
where ${\cal S}_{\phi K_S}$ is the analogous measure in the $B^0
\to \phi K_S$ mode. New physics FCNC and CPV, such as SUSY in the
loop, could break this equality, prompting the experiments to
search vigorously.

Many might remember the big splash made by Belle in 2003, which
found ${\cal S}_{\phi K_S}$ to be opposite in
sign~\cite{phiKs_Belle03} to $\sin2\phi_1/\beta$, deviating by
more than 3$\sigma$. The situation softened by 2004 and is now far
less dramatic, but it has persisted in a nagging way. Comparing to
the average of  ${\cal S}_{c\bar cs} = 0.68 \pm 0.03$~\cite{HFAG}
over $b\to c\bar cs$ transitions, ${\cal S}_f$ is smaller in
practically all $b\to s\bar qq$ modes measured so far, with the
naive mean\footnote{
 We use the EPS2007 result, rather than
 the LP2007 update that includes the new ${\cal S}_{f_0(980)K_S}$
 result from BaBar. Though the latter appears to be larger than
 ${\cal S}_{c\bar cs}$ and very precise, it needs confirmation
 from Belle.}
of ${\cal S}_{s\bar qq} = 0.56 \pm 0.05$~\cite{HFAG}. The
deviation is only 2.1$\sigma$, and has been slowly diminishing. We
stress, however, that the persistence over several years and in
multiple modes make this ``$\Delta{\cal S}$ problem" a potential
indication for New Physics from the B factories, and should be
taken seriously.

The point is that theoretical studies, though troubled by hadronic
effects, all give ${\cal S}_{s\bar qq}$ values above ${\cal
S}_{c\bar cs}$. A model-independent approach suggested~\cite{SMH}
that, with enough precision, a deviation as little as a couple of
degrees would indicate New Physics. Alas, the data can at best
double in the remaining B factory era. Lacking good vertices in
the leading channels of $\eta^\prime K_S$, $\phi K_S$ and $K_S
\pi^0$, the situation may not improve greatly with LHCb. Thus,
this problem would need a Super B factory to clarify.

\subsection{\boldmath $\Delta{{\cal A}_{K\pi}}$}
\label{DeltaAKpi}

There is a second possible indication for BSM in $b\to s\bar qq$.
It is less widely known, but experimentally firm.

Between BaBar and Belle, direct CPV (DCPV) in the B system was
claimed in 2004~\cite{PDG}, just 3 years after the observation of
TCPV in B. This attests to the prowess of the B factories, as it
took 35 years for the same evolution in the K system. The CDF
experiment recently joined the club, with results consistent with
the B factories. The current world average~\cite{HFAG} is ${\cal
A}_{\rm CP}(K^+\pi^-) = -9.7 \pm 1.2\ \%$. This by itself does not
suggest New Physics, but rather, it indicates the presence of a
finite strong phase between the strong penguin (P) and tree (T)
amplitudes, which most QCD based factorization approaches failed
to predict.

Even in 2004, however, there was a hint of a puzzle. In contrast
to the negative value for $B^0 \to K^+\pi^-$, DCPV in the charged
$B^+\to K^+\pi^0$ mode was found to be consistent with zero for
both Belle and BaBar, which has steadily strengthened, to the
current~\cite{HFAG} ${\cal A}_{\rm CP}(K^+\pi^0) = +4.7 \pm 2.6\
\%$, with some significance for the positive sign. This will
further strengthen with an updated value of ${\cal A}_{\rm
CP}(K^+\pi^0) = +3.0 \pm 3.9 \pm 1.0\ \%$ from
BaBar~\cite{AKpi0_BaBar07}. The deviation between the charged and
neutral modes,
\begin{eqnarray}
\Delta{{\cal A}_{K\pi}}
 \equiv {\cal A}_{K^+\pi^0} - {\cal A}_{K^+\pi^-}
 = 0.144 \pm 0.029,
 \label{eq:DeltaAKpi}
\end{eqnarray}
is now beyond $5\sigma$.


\begin{figure}[b!]
\hskip0.35cm
\includegraphics[width=0.43\textwidth,height=0.30\textwidth,angle=0]{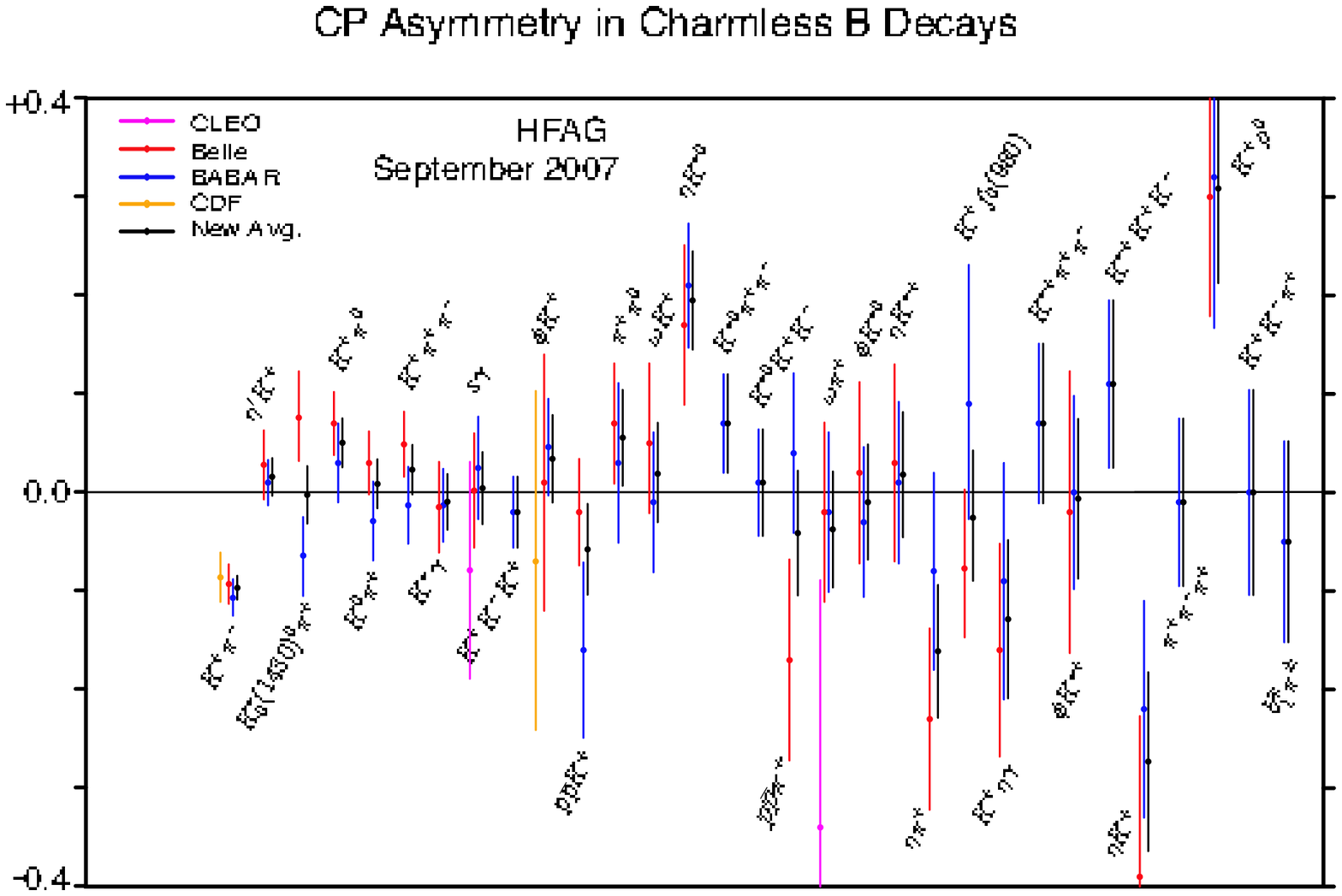}
\vskip0.2cm
 \caption{
 HFAG plot for DCPV measurements.
 The difference between ${\cal A}_{K^+\pi^0}$
 and ${\cal A}_{K^+\pi^-}$, including sign, could indicate New Physics.}
 \label{fig:ACP}       
\end{figure}
%

Why is this a puzzle? For $B^0$ decay mode, one has
\begin{eqnarray}
{\cal M}(B^0 \to K^+\pi^-)
 \propto T + P \equiv r \, e^{i\phi_3} + e^{i\delta},
 \label{eq:MKpi}
\end{eqnarray}
where $r \equiv \vert T/P\vert$, $\phi_3 = \arg V_{ub}^*$, and
$\delta$ is the strong phase difference between $T$ and $P$. It is
the interference between the two kinds of phases that gives DCPV,
i.e. ${\cal A}_{\rm CP}(K^+\pi^-)$.
Note that for TCPV, $\delta = \Delta m_B t$, the measurable
oscillation phase. This is part of the beauty of TCPV.
The $B^+\to K^+\pi^0$ decay amplitude is similar to the $B^0\to
K^+\pi^-$ one, up to subleading corrections,
\begin{eqnarray}
\sqrt{2}{\cal M}_{K^+\pi^0} - {\cal M}_{K^+\pi^-}
 \propto P_{\rm EW} + C,
 \label{eq:DeltaMKpi}
\end{eqnarray}
where $P_{\rm EW}$ is the electroweak penguin (replacing the
virtual gluon in $P$ by $Z$ or $\gamma$) amplitude, and $C$ is the
colour-suppressed tree. In the limit that these subleading terms
vanish, one expects $\Delta{{\cal A}_{K\pi}} \sim 0$, which is
contrary to the experimental result of Eq.~(\ref{eq:DeltaAKpi}).

Could $C$ be greatly enhanced? Indeed, fitting with data, one
finds $\vert C/T\vert > 1$ is needed~\cite{BL07}, in contrast to
the very tiny value 10 years ago~\cite{Neubert98}. Furthermore, as
the amplitude $C$ has the same weak phase $\phi_3$ as $T$, the
enhancement has to contrive to cancel the effect of the strong
phase difference $\delta$ between $T$ and $P$ that helped induce
${\cal A}_{\rm CP}(K^+\pi^-)$, amounting to a ``double
somersault". Next order perturbative QCD calculations do move $C$
in the right direction, but insufficient to account for
Eq.~(\ref{eq:DeltaAKpi}). The SCET approach completely fails in
the DCPV sector.

The other option is to have a sizable contribution from the
electroweak penguin. The interesting point is that {\it this calls
for a New Physics CPV phase}, as it is known that $P_{\rm EW}$ and
$T$ have almost the same weak phase within SM~\cite{NR98}. So what
NP can this be? Note that this is not so easy for SUSY, since SUSY
effects tend to be of the ``decoupling" kind, compared to the {\it
nondecoupling} top effect in the $Z$ penguin loop, which is very
analogous to what happens in box diagrams. So, can there be more
{\it nondecoupled} quarks beyond the top in the $Z$ penguin loop?
We will look further into this, after we discuss NP prospects in
$B_s$ mixing.

With the two hints for NP in $b\to s$ penguin modes, i.e.
$\Delta{\cal S}$ (TCPV) and $\Delta{{\cal A}_{K\pi}}$ (DCPV), one
might expect possible NP in $B_s$ mixing. On the other hand,
recent results for $\Delta m_{B_s}$ and $\Delta \Gamma_{B_s}$ are
SM-like. But the real test should be in the CPV measurables
$\sin2\Phi_{B_s}$ and $\cos2\Phi_{B_s}$, as the NP hints all
involve CPV.

\subsection{\boldmath $B_s$ Mixing and $\sin2\Phi_{B_s}$}
\label{sin2PhiBs}

The oscillation between $B_s^0$ and $\bar B_s^0$ mesons is too
fast for B factories, hence brings us to the hadronic collider
detectors, CDF and D$\emptyset$ at the Tevatron. After a slow
start of the Run II, the experiments have recently reached 3
fb$^{-1}$ integrated luminosity, and is growing steadily.

The special two-track trigger of CDF allowed it to leapfrog the
earlier announcement made by D$\emptyset$ in Winter 2006, and by
Summer 2006, based on 1 fb$^{-1}$, $B_s$ mixing became a precision
measurement~\cite{Bsmix_CDF06},
\begin{eqnarray}
 \Delta m_{B_s} = 17.77 \pm 0.10 \pm 0.07\ {\rm ps}^{-1}.
 \label{eq:DeltamBs}
\end{eqnarray}
We remark that, if one takes the nominal value for $f_{B_s}$ e.g.
from lattice studies, {\it the result of Eq.~(\ref{eq:DeltamBs})
seems a bit on the small side}. Before the fact, the values from
the CKMfitter and UTfit ``global fit" groups tended to be larger
than 20 ps$^{-1}$. However, given the hadronic uncertainty in
$f_{B_s}^2B_{B_s}$, this can hardly be taken as a hint for New
Physics. One has to turn to CPV.

In SM, $M_{12}^s \propto m_t^2 (V_{tb}V_{ts}^*)^2$, and CPV in
$B_s$ mixing is controlled by the phase of $V_{ts}$. Since
$V_{us}^*V_{ub}$ is very tiny, the triangle relation
\begin{eqnarray}
V_{us}^*V_{ub} + V_{cs}^*V_{cb} + V_{ts}^*V_{tb} = 0,
 \label{eq:btosTri}
\end{eqnarray}
collapses to the approximate line of $V_{ts}^*V_{tb} \simeq -
V_{cb}$, which is practically real, or $\arg V_{ts} \sim
-0.02$~rad. Only the LHCb experiment would have enough sensitivity
to probe this. Thus, it is well known that $\sin2\Phi_{B_s}$, the
analogue of $\sin2\phi_1/\beta$ for $B_d$, is an excellent window
on BSM. In SUSY, this could be squark-gluino loop with $\tilde
s$-$\tilde b$ mixing.

\begin{figure}[t!]
\hskip1.2cm
\includegraphics[width=0.35\textwidth,height=0.24\textwidth,angle=0]{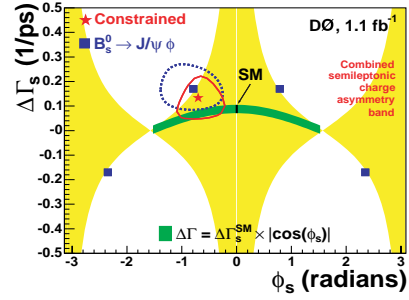}
\caption{
 Combined analysis of $A_{SL}$, $A_{SL}^s$ and lifetime
 difference in $B_s \to J/\psi \phi$ by D$\emptyset$~\cite{cos2PhiBs_Dzero07}.}
 \label{fig:phisDzero}       
\end{figure}

Let us first comment on the approach through width mixing, i.e.
$\Delta\Gamma_{B_s}$ and $\phi_{B_s}$ from $B_s^0 \to J/\psi
\phi$. Here, the D$\emptyset$ experiment has made a concerted
effort on dimuon charge asymmetry $A_{SL}$, the untagged single
muon charge asymmetry $A_{SL}^s$, and the lifetime difference in
untagged $B_s \to J/\psi \phi$ decay (hence does not involve
oscillations). D$\emptyset$ holds the advantage in periodically
flipping magnet polarity to reduce the systematic error on
$A_{SL}$. Combining the three studies, they probe the CPV phase
$\cos2\Phi_{B_s}$ via
\begin{eqnarray}
\Delta \Gamma_{B_s} = \Delta \Gamma_{\rm CP}\, \cos2\Phi_{B_s}.
 \label{eq:cos2PhiBs}
\end{eqnarray}
We give the main result for our interest in
Fig.~\ref{fig:phisDzero}; this ``first" $\cos2\Phi_{B_s}$ value is
slightly off, but consistent with, SM expectation, hence certainly
allows for NP. Details can be found in
Ref.~\cite{cos2PhiBs_Dzero07}.
For a phenomenological digest, see Ref.~\cite{HM07}. Overall,
$\cos2\Phi_{B_s}$ is a somewhat ``blunt instrument".

We return to assessing the short term prospects for $\Phi_{B_s}$
measurement.
$B_s \to J/\psi \phi$ decay is analogous to $B_d \to J/\psi K_s$,
except it is a $VV$ final state. Thus, besides measuring the decay
vertices, one also needs to perform an angular analysis to
separate the $CP$ $+/-$ components.
As $J/\psi$ is reconstructed in say the dimuon final state, CDF
and D$\emptyset$ should have comparable sensitivity. Assuming 8
fb$^{-1}$ per experiment, the Tevatron could reach (?) the
sensitivity of 0.2$/\sqrt{2}$. However, the LHC would start
running a year from now, in 2008. I will adopt a conservative
estimate~\cite{Nakada07} for the first year running of LHC: 2.5
fb$^{-1}$ for ATLAS and CMS, and 0.5 fb$^{-1}$ for LHCb. Then the
projection for ATLAS is $\sigma(\sin2\Phi_{B_s}) \sim 0.16$, not
better than the Tevatron, while for LHCb one has
$\sigma(\sin2\Phi_{B_s}) \sim 0.04$, which starts to probe the SM
expectation of $\sigma(\sin2\Phi_{B_s}) = -0.04$.

\begin{figure}[b!]
\hskip0.9cm
\includegraphics[width=0.40\textwidth,height=0.23\textwidth,angle=0]{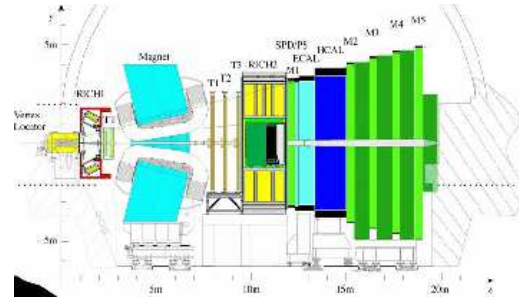}
\vskip0.35cm
 \caption{
 The LHCb detector.}
 \label{fig:LHCb}       
\end{figure}

The forward design of LHCb detector (see Fig.~\ref{fig:LHCb}) is
for $B$ physics, allowing more space for devices such as RICH for
PID.
If SM again holds sway, it would clearly be the winner. We wish to
stress, however, that {\it 2009 looks rather interesting}
--- Tevatron could get really lucky: it could glimpse the value of
$\sin2\Phi_{B_s}$ only if it is large; but {\it if
$\sin2\Phi_{B_s}$ is large, it would definitely indicate New
Physics}. Thus, the Tevatron could preempt LHCb and carry away the
glory.
Maybe the Tevatron should even run beyond 2008? The question is
then,

\vskip0.15cm
 \centerline{\bf\boldmath
             --- \ Can $\vert\sin2\Phi_{B_s}\vert > 0.5$? \ ---}
\vskip0.1cm

%
\begin{table}[t!]
\caption{Rough sensitivity to $\sin2\Phi_{B_s}$ ca. 2009.}
\label{tab:sigma_SBs}       
\begin{tabular}{cccc}
\hline\noalign{\smallskip}
& CDF/D{$\emptyset$} & ATLAS/CMS & LHCb   \\
\noalign{\smallskip}\hline\noalign{\smallskip}
$\sigma(\sin2\Phi_{B_s})$ & 0.2/expt & 0.16/expt & 0.04 \\
$\int {\cal L}dt$ & (8 fb$^{-1}$) & (2.5 fb$^{-1}$) & (0.5 fb$^{-1}$) \\
\noalign{\smallskip}\hline
\end{tabular}
\end{table}

One can of course resort to squark-gluino box diagrams. Note,
however, that squark-gluino loops, while possiblly generating
$\Delta{\cal S}$, cannot really move $\Delta {\cal A}_{K\pi}$
because their effects are decoupled in $P_{\rm EW}$. If one cares
about contact with both hints for NP in $b\to s$ transitions from
the B factories, then one should pay attention to some common
nature between $b\to s$ electroweak penguin and the $B_s$ mixing
box diagram. If there are new {\it nondecoupled} quarks in the
loop, then both $\Delta {\cal A}_{K\pi}$ and $\Delta{\cal S}$
could be touched. Such nondecoupled quarks are traditionally
called the 4th generation. The $t^\prime$ quark in the loop adds a
term $V_{ts}^*V_{tb} \equiv r_{sb}\, e^{i\phi_{sb}}$ to
Eq.~(\ref{eq:btosTri}), bringing in the additional NP CPV phase
$\arg(V_{t's}^*V_{t'b}) \equiv \phi_{sb}$ with {\it Higgs
affinity} $\lambda_{t^\prime} > \lambda_t \simeq 1$.

It was shown~\cite{HNS05} that the 4th generation could affect
$\Delta {\cal A}_{K\pi}$ in the right way, and $\Delta {\cal
S}$~\cite{HLMN07} then moves in the right direction. This was done
in the PQCD approach at NLO, which is state of the art. PQCD is
the only QCD-based factorization approach that predicted both the
strength and sign of ${\cal A}_{\rm CP}(B^0\to K^+\pi^-)$, and at
NLO, saw the improvement of $\Delta {\cal A}_{K\pi}$ by
enhancement of $C$. It is nontrivial, then, that incorporating the
nondecoupled 4th generation $t^\prime$ quark to account for
$\Delta {\cal A}_{K\pi}$, brings $\Delta {\cal S}$ in the right
direction.

The exciting implication is the impact on $\sin2\Phi_{B_s}$. As
the difference of $\Delta {\cal A}_{K\pi}$ in
Eq.~(\ref{eq:DeltaAKpi}) is large, both the strength and phase of
$V_{t's}^*V_{t'b}$ is sizable~\cite{HNS05}, with the phase near
maximal. Interestingly, a near maximal phase for $t^\prime$ allows
minimal impact on $\Delta m_{B_s}$, as it adds only in quadrature
to the real contribution from top, but exerts the maximal impact
on $\sin2\Phi_{B_s}$. The predicted value is~\cite{HNSprd}
\begin{eqnarray}
\sin2\Phi_{B_s} = -0.5\ {\rm to}\ -0.7,
 \ \ \ {\rm (4th\ generation)}
 \label{eq:sin2PhiBs4th}
\end{eqnarray}
where even the sign is predicted. We note that the range can be
demonstrated by using the (stringent) $\Delta m_{B_s}$ vs (less
stringent) ${\cal B}(B\to X_s\ell^+\ell^-)$ constraints alone,
with $\Delta {\cal A}_{K\pi}$ selecting the minus sign in
Eq.~(\ref{eq:sin2PhiBs4th}).

We stress that Eq.~(\ref{eq:sin2PhiBs4th}) can be probed even
before LHCb gets first data, and should help motivate the Tevatron
experiments. {\it It's not over until it's over.}

\subsection{\boldmath ${\cal A}_{\rm CP}(B^+\to J/\psi K^+)$}
 \label{ApsiK+}

Suppose there is New Physics in the $B^+\to K^+\pi^0$ electroweak
penguin. Rather than turning into a $\pi^0$, the $Z^*$ from the
effective $bsZ^*$ vertex could turn into a $J/\psi$ as well. One
can then contemplate DCPV in $B^+\to J/\psi K^+$ as a probe of NP.

$B^+\to J/\psi K^+$ decay is of course dominated by the
colour-suppressed $b\to c\bar cs$ amplitude, which is proportional
to $V_{cs}^*V_{cb}$ and is practically real. At the loop level,
the penguin amplitudes are proportional to $V_{ts}^*V_{tb}$.
Because $V_{us}^*V_{ub}$ is very suppressed, from
Eq.~(\ref{eq:btosTri}) one has $V_{ts}^*V_{tb} \cong -
V_{cs}^*V_{cb}$. It is not only practically real, but has the same
phase as the tree amplitude, hence it is commonly argued that DCPV
is less than $10^{-3}$ in this mode. However, because of possible
hadronic effects, there is no firm prediction that can stand
scrutiny.
We shall argue that, in the 4th generation scenario, DCPV in
$B^+\to J/\psi K^+$ decay could be at $\%$ level.

Experiment so far is consistent with zero, but has a somewhat
checkered history. Belle has not updated from their 2003 study
based on 32M $B\bar B$ pairs, although they now have almost
20$\times$ the data. BaBar's study flipped sign from the 2004
study based on 89M, to the 2005 study based on 124M, which seemed
dubious at best. However, the sign was flipped back in PDG 2007,
because it was found that the 2005 paper simply used the opposite
convention to the (standard) one used for 2004. The opposite sign
between Belle and BaBar suppresses the central value, but the
error is at 2\% level. This rules out, for example, the
suggestion~\cite{WS00} of enhanced $H^+$ effect at 10\% level.

One impediment to higher statistics B factory studies is the
systematic error, where it seems difficult to break the 1\%
barrier.
Recent progress has been made, however, by D$\emptyset$. Using 1.6
fb$^{-1}$, D$\emptyset$ measures~\cite{ApsiK+_Dzero07}
\begin{eqnarray}
{\cal A}_{B^+\to J/\psi K^+}
 = 0.67 \pm 0.74 \pm 0.26\ \%, \ \ \ ({\rm D}\emptyset)
 \label{eq:ApsiK+}
\end{eqnarray}
where there is a large correction for the $K^\pm$ asymmetry of the
detector due to matter effect. Of special note is the small (below
0.5\% level!) systematic error. This is because one has a larger
control sample than at B factories, e.g. in $D^*$ tagged $D^0\to
K^-\pi^+$ decays. Thus, even scaling up to 8 fb$^{-1}$, one is
still statistics limited, and one would have 2$\sigma$ sensitivity
with \% level asymmetries. CDF should have similar sensitivity,
and the situation can drastically improve with LHCb.

The Tevatron study was in fact inspired by a 4th generation
study~\cite{HNSpsiK+} following the lines of the previous
sections. The 4th generation parameters are taken from the $\Delta
{\cal A}_{K\pi}$ study. By analogy with what is observed in $B \to
D\pi$ modes, as well as between different helicity components in
$B\to J/\psi K^*$ decay, the dominant $C$ amplitude for $B^+\to
J/\psi K^+$ would likely have a strong phase of order $30^\circ$.
The $P_{\rm EW}$ amplitude is assumed to factorize and hence does
not pick up a strong phase. Heuristically this is because the
$Z^*$ produces a small, colour singlet $c\bar c$ that projects
into $J/\psi$. With a strong phase in $C$ and a weak phase in
$P_{\rm EW}$, one then finds ${\cal A}_{B^+\to J/\psi K^+} \simeq
\pm 1\%$, with negative sign ruled out by Eq.~(\ref{eq:ApsiK+}).
Of course, DCPV is directly proportional to the strong phase
difference, which is not well predicted. However, if \% level
asymmetry is observed in the next few years, it would support the
scenario of New Physics in $b\to s$ transitions, while stimulating
theoretical efforts to compute the strong phase difference between
$C$ and $P_{\rm EW}$.

\section{\boldmath $H^+$ Probes}
\label{sec:H+}
\subsection{\boldmath $b\to s\gamma$}
\label{sec:btosgamma}

The inclusive $b\to s\gamma$ decay, identified with $B\to
X_s\gamma$ experimentally, where $X_s$ are reconstructed as
$K+n\pi$, is one of the most important probes of NP. There had
been good agreement for the past few years between NLO theory and
the experimental average of
\begin{eqnarray}
{\cal B}_{B\to X_s\gamma} = (3.55\pm0.26) \times 10^{-4}\
 {\rm (HFAG\;06)}.
 \label{eq:Xsgamma}
\end{eqnarray}

Recently, however, the NNLO theory prediction has shifted
lower~\cite{Misiak07,BN07} to $\sim 3 \times 10^{-4}$, with errors
comparable to experiment. Although the NNLO work is not yet
complete, the ball appears to be in the experiments' court.

\begin{figure}[t!]
\hskip1.75cm
\includegraphics[width=0.28\textwidth,height=0.26\textwidth,angle=0]{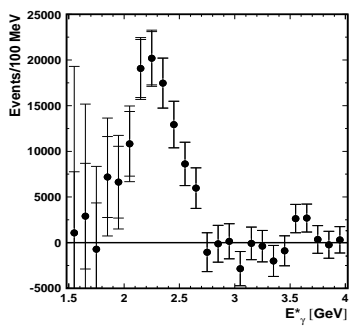}
 \vskip0.5cm
\caption{
 The $E_\gamma$ spectrum above 1.8 GeV in $\Upsilon(4S)$ frame
 for inclusive $b\to s\gamma$
 (Belle~\cite{bsgamma_Belle04} 152M $B\bar B$ pairs).}
 \label{fig:Egamma}       
\end{figure}

The photon energy cut, where the latest Belle study sets at
$E_\gamma > 1.8$ GeV~\cite{bsgamma_Belle04} (see
Fig.~\ref{fig:Egamma}) and done for 152M $B\bar B$ pairs, should
be lowered further. But, to confront the theoretical advancement,
a fresher approach is needed. For example, the $X_s \equiv K+n\pi$
``partial reconstruction" technique is over a decade old. A
promising new development, as the B factories increase in data, is
the full reconstruction of the tag side $B$ meson. The signal side
is then just an energetic photon, without specifying the $X_s$
system. First attempts have recently been performed by BaBar, but
since full reconstruction takes a $10^{-3}$ hit in efficiency, it
seems that the NNLO theory development would demand a Super B
factory upgrade to continue the supreme dialogue between theory
vs. experiment in this mode.

This close dialogue allowed $b\to s\gamma$ to provide one of the
most stringent bounds on NP models. It is sensitive to all types
of possible NP in the loop, such as stop-charginos. However, $b\to
s\gamma$ is best known for its stringent constraint on the MSSM
(minimal SUSY SM) type of $H^+$. MSSM demands at least two Higgs
doublets (2HDM), where one Higgs couples to right-handed down type
quarks, the other to up type. The physical $H^+$ is a cousin of
the $\phi_{W^+}$ Goldstone boson of the SM
that gets eaten by the $W^+$. It is the $\phi_W$ that couples to
masses, and at the root of the nondecoupling phenomenon of the
heavy top quark in the loop. In $bs\gamma$ coupling, however, the
top is effectively decoupled, by a subtlety of gauge invariance.
This underlies the reason why QCD corrections make such large
impact in this loop-induced decay. It also makes the process
sensitive to NP such as $H^+$.

Replacing the $W^+$ by $H^+$ in the loop, in the MSSM type of
2HDM, the $H^+$ effect {\it always enhances $b\to s\gamma$ rate,
regardless of $\tan\beta$}, as pointed out 20 years
ago~\cite{GW88,HW88}, where $\tan\beta$ is the ratio of v.e.v.s
between the two doublets. Basically, the $H^+$ couples to
$m_t\cot\beta $ at one end of the loop, and to $-m_b\tan\beta$ on
the other end, so this contribution is independent of $\tan\beta$,
and the sign is fixed to be always constructive with the SM
amplitude. With NNLO result lower than experiment, one has the
bound~\cite{Misiak07}
\begin{eqnarray}
m_{H^+} > 295\ {\rm GeV\ (90\% C.L.)},
 \label{eq:H+bound}
\end{eqnarray}
if one takes the low range of NNLO result and compare with the
higher range of Eq.~(\ref{eq:Xsgamma}). A nominal $\tan\beta = 2$
is taken. If one takes the central value of both results
seriously, one could say~\cite{Misiak07} that an $H^+$ boson with
mass around 695 GeV is needed to bring the NNLO rate up to
Eq.~(\ref{eq:Xsgamma}). Again, this is because the $H^+$ effect in
the MSSM type of 2HDM is always constructive~\cite{HW88} with
$\phi_W$ effect in SM.
We note that the behavior for the other 2HDM that is not the MSSM
type, the $H^+$ effect is different~\cite{HW88}.

The ongoing saga should be watched.

\subsection{\boldmath $B\to (D^{(*)})\tau\nu$}
\label{sec:btotaunu}

As a cousin of the $\phi_{W^+}$, the $H^+$ boson has an amazing
tree level effect that has only recently come to fore by the
prowess of the B factories.

Like $\pi^+\to \ell^+\nu_\ell$ decay, one has the formula for
$B^+\to \tau^+\nu_\tau$ decay,
\begin{eqnarray}
 {\cal B}_{B\to \tau\nu} 
 = r_H \frac{G_F^2 m_B m_\tau^2}{8\pi}
       \left[1-\frac{m_\tau^2}{m_B^2}\right]
       \tau_{B} f_B^2\vert V_{ub}\vert^2,
 \label{eq:Btaunu}
\end{eqnarray}
where $r_H = 1$ for SM, but~\cite{Hou93}
\begin{eqnarray}
 r_H = \left[1-\frac{m_{B^+}^2}{m_{H^+}^2}\tan^2\beta\right]^2,
 \label{eq:rH}
\end{eqnarray}
for 2HDM.
Within SM, the pure gauge $W^+$ effect is helicity suppressed,
hence the effect vanishes with the $m_\tau$ mass. For $H^+$, there
is no helicity suppression, but one has the ``Higgs affinity"
factor, i.e. mass dependent couplings. With $m_u$ negligible, the
$H^+$ couples as $m_\tau m_b \tan^2\beta$. This leads to the $r_H$
factor of Eq.~(\ref{eq:rH}), where the sign between the SM and
$H^+$ contribution is always destructive~\cite{Hou93}.

\begin{figure}[t!]
\hskip1.9cm
\includegraphics[width=0.30\textwidth,height=0.21\textwidth,angle=0]{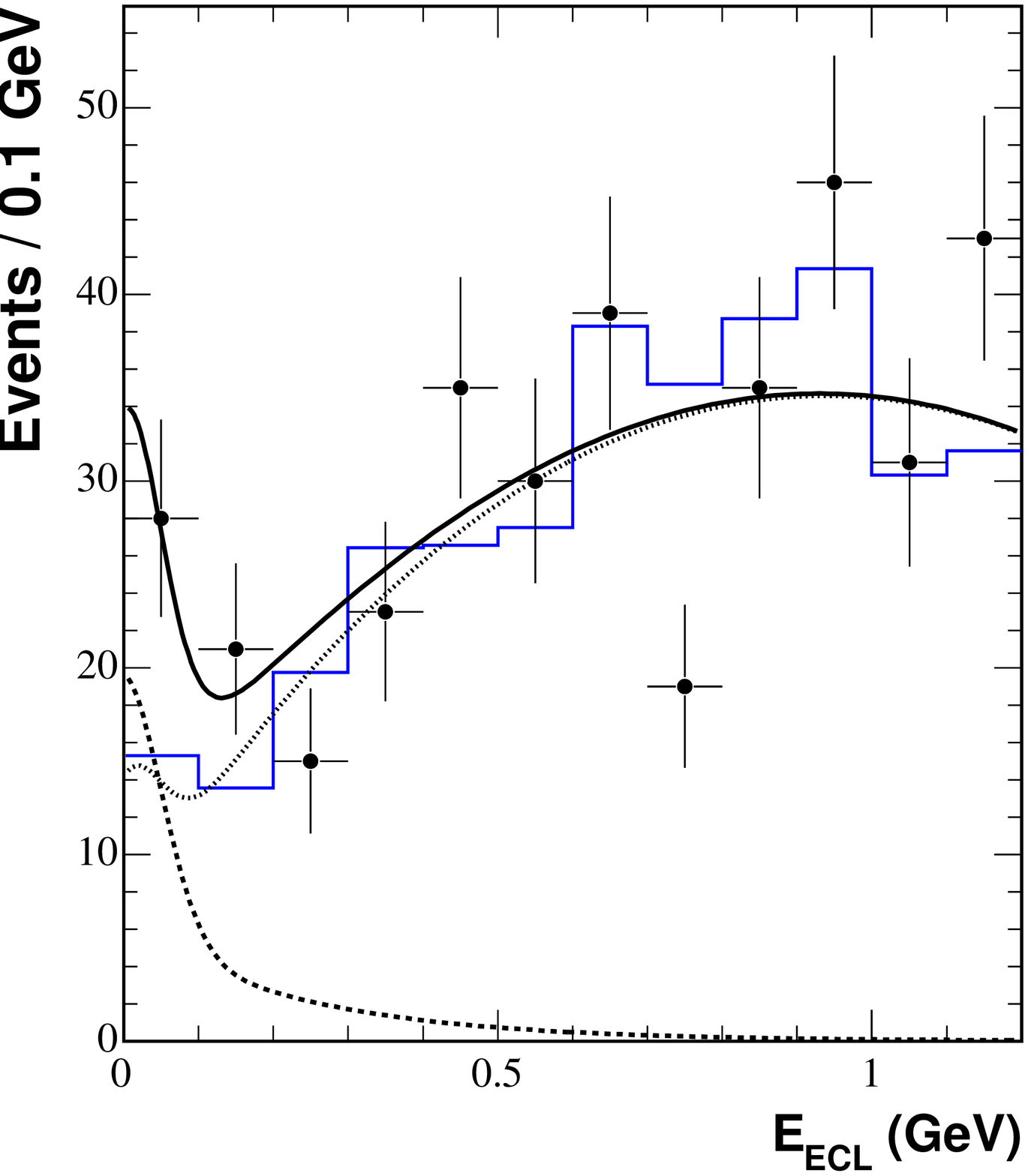}
\vskip-0.07cm\hskip1.57cm
\includegraphics[width=0.315\textwidth,height=0.22\textwidth,angle=0]{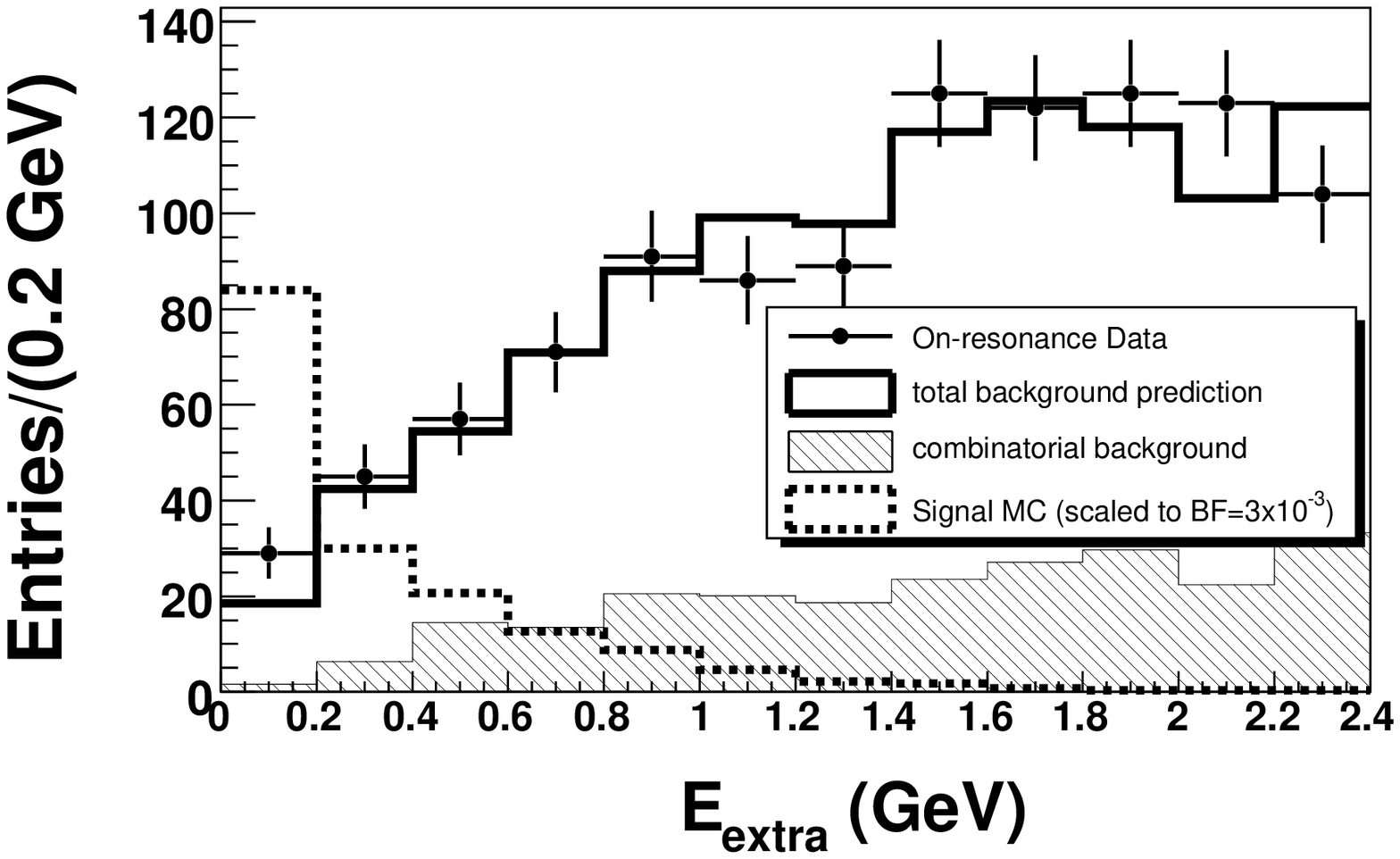}
\vskip0.2cm
\caption{
 Data showing evidence for $B\to\tau\nu$ (hadronic tag) search
 by Belle~\cite{taunu_Belle06} and BaBar~\cite{taunu_BaBar07}.}
 \label{fig:taunu}       
\end{figure}

$B^+\to \tau^+\nu$ followed by $\tau^+$ decay results in at least
two neutrinos, which makes backgrounds very hard to suppress in
the $B\bar B$ production environment. Thus, for a long time, the
limit on $B^+\to \tau^+\nu$ was rather poor. This had allowed for
the possibility that $H^+$ effect could be the dominant one over
SM, given that the SM expectation was only at $10^{-4}$ level. The
change came with the enormous number of B mesons accumulated by
the B factories, allowing the aforementioned full reconstruction
method to become useful.

Fully reconstructing the tag side B meson in, e.g. $B^-\to
D^0\pi^-$ decay, one has an efficiency of only 0.1\%--0.3\%. At
this cost, however, one effectively has a ``B beam".
As shown in Fig.~\ref{fig:taunu}, using full reconstruction in
hadronic modes and with a data consisting of 449M $B\bar B$ pairs,
Belle found $17.2^{+5.3}_{-4.7}$ events, where the $\tau$ decay
was searched for in decays to $e\nu\nu$, $\mu\nu\nu$, $\pi\nu$ and
$\rho\nu$ modes. This constituted the first evidence (at
3.5$\sigma$) for $B^+\to \tau^+\nu$, with~\cite{taunu_Belle06}
\begin{eqnarray}
  {\cal B}_{B\to \tau\nu}  = 1.79^{+0.56+0.46}_{-0.49-0.51} \times 10^{-4}\
 {\rm (Belle\;449M)}.
 \label{eq:taunuBelle}
\end{eqnarray}
With 320M $B\bar B$s and $D\ell\nu$ reconstruction on tag side,
however, BaBar saw no clear signal, giving
$(0.88^{+0.68}_{-0.67}\pm0.11) \times 10^{-4}$.
Updating more recently to 383M, BaBar finds with $D\ell\nu$ tag
the result $(0.9\pm0.6\pm0.1) \times 10^{-4}$, which is consistent
with 320M result. However, with hadronic tag, BaBar now also
reports some evidence, at $(1.8^{+0.9}_{-0.8}\pm0.4\pm0.2) \times
10^{-4}$ (Fig.~\ref{fig:taunu}). The combined result for BaBar
is~\cite{taunu_BaBar07},
\begin{eqnarray}
  {\cal B}_{B\to \tau\nu}
   = (1.2\pm0.4_{\rm stat}\pm0.3_{\rm bkg}\pm0.2_{\rm eff}) \times
   10^{-4}\nonumber \\
 {\rm (BaBar\;383M)},\ \
 \label{eq:taunuBaBar}
\end{eqnarray}
which has 2.6$\sigma$ significance (``bkg" stands for background
and ``eff" stands for efficiency), and is more or less consistent
with the Belle result.

%

Taking central values from lattice for $f_B$, and $\vert
V_{ub}\vert$ from semileptonic decays, the nominal SM expectation
is $(1.6\pm0.4) \times 10^{-4}$. Thus, Belle and BaBar have
reached SM sensitivity, and Eqs.~(\ref{eq:taunuBelle}) and
(\ref{eq:taunuBaBar}) now place a constraint on the
$\tan\beta$-$m_{H^+}$ plane through $r_H \simeq 1$.
If one has a Super B factory, together with development of lattice
QCD, this can become a superb probe of the $H^+$, complementary to
direct $H^+$ searches at the LHC.

An analogous mode with larger branching ratio, $B\to
D^{(*)}\tau\nu$, has recently emerged. Belle announced the
observation of~\cite{Dstartaunu_Belle07}
\begin{eqnarray}
  {\cal B}_{D^{*-}\tau\nu} = 2.02^{+0.40}_{-0.37}\pm0.37\; \%\;\
 {\rm (Belle\;535M)},
 \label{eq:DstartaunuBelle}
\end{eqnarray}
based on $60^{+12}_{-11}$ reconstructed signal events, which is a
5.2$\sigma$ effect. Subsequently, based on 232M, BaBar announced
the observation (over 6$\sigma$) of $D^{*0}\tau\nu$, and evidence
(over 3$\sigma$) for $D^{+}\tau\nu$~\cite{Dstartaunu_BaBar07}
\begin{eqnarray}
  {\cal B}_{D^{*}\tau\nu} &=& 1.81\pm0.33\pm0.11\pm0.06\; \% \nonumber \\
  {\cal B}_{D\tau\nu} &=& 0.90\pm0.26\pm0.11\pm0.06\; \%\;\
 \label{eq:DstartaunuBaBar} \\
 &&{\rm \hskip2cm (BaBar\;232M)}, \nonumber
 \label{eq:DstartaunuBaBar}
\end{eqnarray}
where the last error is from normalization.

The SM branching ratios, at 1.4\% for $B\to D^{*}\tau\nu$, are
poorly estimated. Furthermore, though the $H^+$ could hardly
affect the $B \to D^{*}\tau\nu$ rate, it could leave its mark on
the $D^*$ polarization. The $B \to D\tau\nu$ rate, like $B\to
\tau\nu$ itself, is more directly sensitive to $H^+$~\cite{GH}.
More theoretical work, as well as polarization information, would
be needed for BSM (in particular, $H^+$ effect) interpretation.
But it is rather curious that, almost 25 years after the first B
meson was reconstructed, we have a newly measured mode with $\sim$
2\% branching faction!

\section{\boldmath Electroweak Penguin: $Z$-loop, $Z^\prime$, DM}
 \label{sec:EWP}
\subsection{\boldmath $A_{\rm FB}(B\to K^*\ell\ell)$}
\label{sec:AFB}

The $B\to K^*\ell^+\ell^-$ process ($b\to s\ell^+\ell^-$
inclusively) arises from photonic penguin, $Z$ penguin and box
diagrams. The top quark exhibits nondecoupling in the latter
diagrams, analogous to the box diagrams for $M^0$-$\bar M^0$
mixing. It turns out that the $Z$ penguin dominates the $b\to
s\ell^+\ell^-$ decay amplitude~\cite{HWS87}. Interference between
the vector ($\gamma$ and $Z$) and axial vector ($Z$ only)
contributions to $\ell^+\ell^-$ production gives rise to an
interesting forward-backward asymmetry~\cite{AFB91}. This is akin
to the familiar ${\cal A}_{\rm FB}$ in $e^+e^-\to f\bar f$, except
the enhancement of $bsZ$ penguin brings the $Z$ much closer to the
$\gamma$ in $B$ decay, and one probes potential New Physics in the
loops.

Both inclusive $B\to X_s\ell^+\ell^-$ and exclusive $B\to
K^{(*)}\ell^+\ell^-$ decays have now been measured~\cite{PDG}, and
interest has turned to ${\cal A}_{\rm FB}$ for $B\to
K^{*}\ell^+\ell^-$,
\begin{eqnarray}
  {\cal A}_{\rm FB}(q^2) &=& -C_{10} \xi(q^2)\left[{\rm Re}(C_9)F_1 +
  \frac{1}{q^2}C_7 F_2\right],
 \label{eq:AFB}
\end{eqnarray}
where $C_{i}$ are Wilson coefficients, and formulas for $\xi(q^2)$
and the form factor related functions $F_1$ and $F_2$ can be found
in Ref.~\cite{AFB00}. The study by Belle with 386M $B\bar B$
pairs~\cite{AFB_Belle06} is consistent with SM, and rules out the
possibility of flipping the sign of $C_9$ or $C_{10}$ separately
from SM value, but having both $C_9$ or $C_{10}$ flipped in sign
is not ruled out. BaBar took the more conservative approach of
giving ${\cal A}_{\rm FB}$ in just two $q^2$ bins, below and above
$m_{J/\psi}^2$. With 229M, the higher $q^2$ bin is
consistent~\cite{AFB_BaBar06} with SM and disfavors BSM scenarios.
Interestingly, in the lower $q^2$ bin, while sign-flipped BSM's
are less favored, the measurement is $\sim 2\sigma$ away from SM.
Unfortunately, statistics are poor, which cannot be much improved
without a Super B factory. But this is a domain where LHCb can do
very well.

In the context of LHCb prospects, it was recently
noticed~\cite{HHM07} that, in Eq.~(\ref{eq:AFB}), there is no
reason {\it a priori} to keep the Wilson coefficients real when
probing BSM physics! Note that ${\rm Re}(C_9)$ in
Eq.~(\ref{eq:AFB}) differs from $C_9$ within SM by just a small
correction arising from long distance $c\bar c$ effects. But if
one keeps an open mind (rather than, for example, taking the
oftentimes tacitly assumed Minimal Flavour Conservation mindset),
Eq.~(\ref{eq:AFB}) should be replaced by
\begin{eqnarray}
   {\rm Re}\left(C_9^{\rm eff}C_{10}^*\right) {\cal F}_1 
   + \frac{1}{q^2}{\rm Re}\left(C_7^{\rm eff}C_{10}^*\right) {\cal F}_2,
 \label{eq:AFBcomplex}
\end{eqnarray}
where ${\cal F}_i$ are form factor combinations.
Eq.~(\ref{eq:AFBcomplex}) can exhibit a richer interference
pattern than Eq.~(\ref{eq:AFB}).

\begin{figure}[t!]
\hskip1.5cm
\includegraphics[width=0.30\textwidth,height=0.21\textwidth,angle=0]{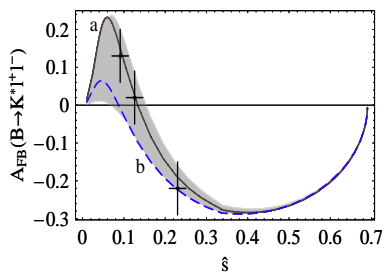}
 \vskip0.2cm
\caption{
 Possible ${\cal A}_{\rm FB}$ in $B\to K^{*}\ell^+\ell^-$ allowed
 by complex Wilson coefficients, Eq.~(\ref{eq:AFBcomplex}). The three data
 points are taken from 2 fb$^{-1}$ LHCb Monte Carlo for illustration.}
 \label{fig:AFB_NP}       
\end{figure}

We illustrate~\cite{HHM07} in Fig.~\ref{fig:AFB_NP} the situation
where New Physics enters through effective $bsZ$ and $bs\gamma$
couplings. In this case, $C_9$ and $C_{10}$ cannot differ by much
at short distance, which gives rise to the ``degenerate tail" for
larger $\hat s \equiv q^2/m_B^2$. But allowing the Wilson
coefficients to be only constrained by measured radiative and
electroweak penguin rates, ${\cal A}_{\rm FB}$ could vary in the
shaded region, basically for $q^2 < m_{J/\psi}^2$, and not just in
the position of the zero. The fourth generation with parameters as
determined from $\Delta m_{B_s}$, ${\cal B}(B\to X_s\ell^+\ell^-)$
and $\Delta {\cal A}_{K\pi}$ belongs to this class of BSM models,
and is plotted as the dashed line. We take the MC study for 2
fb$^{-1}$ data at LHCb (achievable in a couple years of running)
and plot three sample data points to illustrate expected data
quality.

It is clear that the LHCb has good discovery potential using
${\cal A}_{\rm FB}$ to probe complexity of short distance Wilson
coefficients without measuring CPV. If there are New Physics that
affects the $bs\ell\ell$ as a 4-quark operator, for example in
$Z^\prime$ models with FCNC couplings, the allowed range for
${\cal A}_{\rm FB}$ is practically unlimited. If such large
effects are uncovered, one would expect sizable CPV in $b\to
s\gamma$~\cite{HHM07}.

\subsection{\boldmath $B\to K^{*}\nu\nu$}
\label{sec:Knunu}

The $B\to K^*\nu\nu$ (and $b\to s\nu\nu$) is attractive from the
theory point of view that it can arise only from short distance
physics, such as $Z$ penguin and box diagram
contributions~\cite{HWS87}. The photonic penguin does not
contribute. In turn, these processes allow us to probe, in
principle, what happens in the loop. Interestingly, since the
neutrinos go undetected, the process also allows us to probe light
dark matter (DM), which is complementary to the DAMA/CDMS type of
direct search. For instance, DM pairs could arise from exotic
Higgs couplings to the $b\to s$ loop. BaBar has pioneered $B\to
K^{*}\nu\nu$ search. More recently Belle has searched in many
modes with a large dataset of 535M $B\bar B$
pairs~\cite{Knunu_Belle07}, using the aforementioned method of
fully reconstructing the other $B$. No signal is found, and the
most stringent limit is $1.4 \times 10^{-5}$ in $B^+\to
K^+\nu\nu$. While this is still a factor of 3 above SM
expectation, it strengthens a bound on light DM production in
$b\to s$ transitions~\cite{Bird04}.

To measure the theoretically clean $B\to K^{*}\nu\nu$ modes, one
again requires a Super B factory, and there is no resort to LHCb.

We remark, in this context, that Belle has made a special data run
on the $\Upsilon(3S)$ to pursue DM search via $\Upsilon(3S) \to
\pi^+\pi^-\Upsilon(1S)$ followed by $\Upsilon(1S) \to nothing$,
with $\pi^+\pi^-$ and $\Upsilon(3S)$ kinematics as tag. No signal
is found, and only a limit is set~\cite{3SDM_Belle07}.

\section{\boldmath RH Currents and Scalar Interactions}
\label{sec:RHscalar}

\subsection{\boldmath TCPV in $B\to X_0\gamma$}
\label{sec:SX0gamma}

It is clear that the SM, with large QCD enhancement, dominates the
$b\to s\gamma$ rate. The left-handedness of the weak interaction
dictates that the $\gamma$ emitted in $\bar B^0\to \bar
K^{*0}\gamma$ decay has left-handed helicity (defined somewhat
loosely), with the emission of right-handed photons suppressed by
$\sim m_s/m_b$. This reflects the need for a mass insertion for
helicity flip, and the fact that a power of $m_b$ is needed for
the $b\to s\gamma$ vertex by gauge invariance (or current
conservation). For $B^0\to K^{*0}\gamma$ decay that involves $\bar
b\to \bar s\gamma$, the opposite is true, and the emitted photon
is dominantly of RH kind.

An interesting insight can then be made. Mixing-dependent CPV,
i.e. TCPV, involves the interference of $\bar B^0\to \bar
K^{*0}\gamma$ direct decay with $\bar B^0 \stackrel{\rm
mix}{\Longrightarrow} B^0\to K^{*0}\gamma$ decay. The former
process produces $\gamma_L$ while the latter $\gamma_R$, which are
orthogonal hence cannot interfere! The interference is suppressed
by $m_s/m_b \sim$ few \% within SM. However, if there are RH
interactions that induce $b\to s\gamma$ transitions, then $\bar
B^0\to \bar K^{*0}\gamma$ would also have a $\gamma_R$ component
to interfere with the $\bar B^0 \Rightarrow B^0\to K^{*0}\gamma$
amplitude. Thus, TCPV in $B^0\to K^{*0}\gamma$ decay mode
probes~\cite{AGS97} RH interactions!

Alas, Nature plays a trick on us: $K^{*0}\gamma$ has to be in a
$CP$ eigenstate. This means that one needs $K^{*0} \to
K_S^0\pi^0$, and the final state is $K_S^0\pi^0\gamma$. The $K_S$
typically decays at the edge of the silicon detector, and one has
poor vertex information. Fortunately, BaBar
demonstrated~\cite{SKspi0gam_BaBar04} that ``$K_S$ vertexing",
though degraded, was still possible.

The current status of TCPV in $B^0\to K^{*0}\gamma$ decay,
combining the 535M $B\bar B$ pair result from
Belle~\cite{Kstargamma_Belle06}, and the 232M result from
BaBar~\cite{Kstargamma_BaBar05}, gives the average of ${\cal
S}_{K_S\pi^0\gamma} = -0.28\pm 0.26$, which is consistent with
zero. A recent BaBar update with 431M
gives~\cite{Kstargamma_BaBar07} ${\cal S}_{K_S\pi^0\gamma} =
-0.08\pm 0.31\pm0.05$. Measurements have also been made in
$K_s\pi^0\gamma$ mode without requiring $K_s\pi^0$ reconstruct to
a $K^*$.

Again, a Super B factory is needed to probe further, but this is a
very interesting direction to explore.
Other ideas to probe RH currents in $b\to s\gamma$ are $\gamma\to
e^+e^-$ conversion, $\Lambda$ polarization in $\Lambda_b \to
\Lambda\gamma$ decay, and angular $F_L$ and $A_T$ measurables in
$B\to K^*\ell^+\ell^-$.

\subsection{\boldmath $B_s\to \mu\mu$}
\label{sec:Bsmumu}

$B_s\to \mu^+\mu^-$ decay has been a favorite mode to probe Higgs
sector effects in MSSM, because of possible large $\tan\beta$
enhancement.

The process proceeds in SM just like $b\to s\ell^+\ell^-$, except
$\bar s$ is the spectator quark that annihilates the $b$ quark.
Since $B_s$ is a pseudoscalar, the photonic penguin does not
contribute, and one is sensitive to scalar operators. The SM
expectation is only at the $3.5\times 10^{-9}$ level, because of
helicity suppression. In MSSM, a $t$-$W$-$H^+$ loop can emit
neutral Higgs bosons that turn into muon pairs, giving rise to an
amplitude $\propto \tan^6\beta$~\cite{BK00}, which could greatly
enhance the rate even with modest pseudoscalar mass $m_A$.
Together with the ease for trigger and the enormous number of $B$
mesons produced, this is the subject vigorously pursued at hadron
facilities, where there is enormous range for search.

With Run-II data taking shape, the Tevatron experiments have
improved the limits on this mode considerably. The recent 2
fb$^{-1}$ limits from CDF and D$\emptyset$ are $< 5.8 \times
10^{-8}$~\cite{BsmumuCDF07} and $9.3\times
10^{-8}$~\cite{BsmumuDzero07} respectively, combining to give
${\cal B}(B_s\to \mu^+\mu^-) < 4.5 \times 10^{-8}$. This is still
an order of magnitude away from SM.

The expected reach for the Tevatron is $2 \times 10^{-8}$. Further
improvement would have to come from LHCb. LHCb claims that, with
just 0.05 fb$^{-1}$, it would overtake the Tevatron, attain
3$\sigma$ evidence for SM signal with 2 fb$^{-1}$, and 5$\sigma$
observation with 10 fb$^{-1}$. To follow our modest 0.5 fb$^{-1}$
expectation for the first year of LHCb data taking, we expect LHCb
to exclude branching ratio values down to SM expectation.

Clearly, much progress will come with the turning on of LHC, where
direct search for Higgs particles and charginos would also be
vigorously pursued.

\section{\boldmath D/K: Box and EWP Redux}
\label{sec:D/K}

We touch upon $D$ and $K$ mesons only very briefly.

\subsection{\boldmath $D^0$ Mixing}
\label{sec:Dmix}

$D^0$-$\bar D^0$ mixing is the only neutral meson mixing yet to be
observed. In 2007, it was claimed. This is quite some feat of
experiment.

Box diagrams, much like $K^0$, $B_d^0$ and $B_s^0$ meson system,
govern short distance contributions to $D^0$ mixing.
Unfortunately, the $d$ and $s$ quark masses are small compared to
$m_b$ (which is also tiny compared to $m_t$), hence only $b$ quark
contributes in the box at short distance. But $V_{ub}V_{cb}^*$ is
extremely small compared to the leading $V_{ud}V_{cd}^*\simeq -
V_{us}V_{cs}^* \cong -0.22$ in the CKM triangle relation
$V_{ud}V_{cd}^*+V_{us}V_{cs}^*+V_{ub}V_{cb}^*$. Thus, the short
distance contribution to $D^0$ mixing is very small, making it
susceptible to long distance contributions. It has been argued
that the latter can generate a percent level {\it width
difference}, $y_D = \Delta\Gamma_D/2\Gamma_D$. In turn, a
$\Delta\Gamma_D$ at the percent level can generate~\cite{Falk04},
via dispersion, a comparable width mixing $x_D = \Delta
m_D/\Gamma_D$. Unfortunately, so far this seems to be what is
observed.

\begin{figure}[t!]
\hskip1.2cm
\includegraphics[width=0.35\textwidth,height=0.24\textwidth,angle=0]{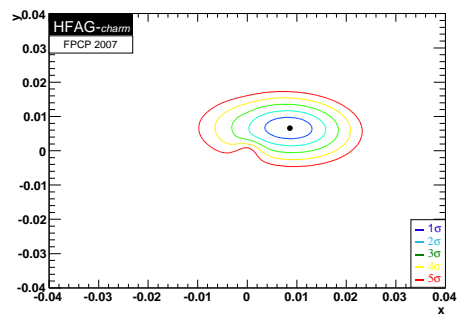}
 \vskip0.8cm
\caption{
 HFAG plot of combined fit to $D^0$ mixing data,
 with Eq.~(\ref{eq:Dmixing}) as best fit result,
 together with $\delta_D = 0.33^{+0.26}_{-0.29}$.}
 \label{fig:Dmixing}       
\end{figure}

Belle has analyzed 540 fb$^{-1}$ data in $D^0\to K^+K^-$,
$\pi^+\pi^-$ ($CP$ eigenstates) to extract $y_{CP}$, and a Dalitz
analysis of $D^0\to K_S\pi^+\pi^-$ to extract $x_D$ and $y_D$.
Both Belle and BaBar have analyzed $D^0\to K^\mp\pi^\pm$ (Cabibbo
allowed vs suppressed), with 384 fb$^{-1}$ and 400 fb$^{-1}$ data
respectively, to extract $x_D^{\prime 2}$ and $y_D^\prime$, where
$x_D^{\prime}$ and $y_D^\prime$ is a rotation from $x_D$ and $y_D$
by a strong phase $\delta_D$ between the Cabibbo allowed and
suppressed $D^0\to K^\mp\pi^\pm$ decays. The analyses are too
complicated to report here. Suffice it to say that currently
$(x_d,\; y_D) = (0,\; 0)$ is excluded at the 5$\sigma$ level (see
Fig.~\ref{fig:Dmixing}), and $D^0$ mixing is now observed. The
current best fit, assuming $CP$ invariance, gives,
\begin{eqnarray}
   x_D = 0.87^{+0.30}_{-0.34}\,\%, \ 
   y_D = 0.66^{+0.21}_{-0.20}\,\%, \ 
 \label{eq:Dmixing}
\end{eqnarray}
with $\delta_D = 0.33^{+0.26}_{-0.29}$. While $y_D$ is more solid,
a finite \% level $x_D$ is indicated. Although the observed
strength could arise from long distance effects, one recalls the
$\Delta m_K$ enterprise 20-30 years ago: comparable BSM, at twice
the observed $x_D$, is always allowed.

For the future, there are several things to watch. It is
interesting that the Dalitz analysis of
Belle~\cite{D0Kspipi_Belle07} sees, for the first time, an
indication for $x_D$. Second, by a tagged Dalitz analysis in
$\psi(3770) \to D^0\bar D^0$, one can extract the phase
$\delta_D$, which would in turn feedback on $x$ and $y$
extraction. Here, CLEO-c and BES-II can contribute. These are the
things to watch. Ultimately, one would need to measure CPV,
expected to be tiny within SM (with or without long distance
dominance), to find unequivocal evidence for BSM.

This is an area where a Super B factory can compete well with
LHCb.

\subsection{\boldmath Rare $K$}
\label{sec:rareK}

This field saw its last hurrah in
$\varepsilon^\prime/\varepsilon$. Unfortunately, the
interpretation of $\varepsilon^\prime/\varepsilon$ is almost
completely clouded by long-distance effects.

With the cancellations of CKM and KOPIO, the kaon program in the
US has withered, despite a long standing hint of 3 events for
$K^+\to \pi^+\nu\nu$ at BNL by E787/949. At CERN, one now has the
P236 proposal to use the SPS, aiming at reaching ${\cal O}(100)$
events with the SM branching ratio of $\sim 10^{-10}$. In Japan,
one has the on-going E391A experiment at KEK. The expected reach
for $K_L\to \pi^0\nu\nu$ is $10^{-9}$, not sufficient to probe the
SM expectation of $10^{-11}$, although there is New Physics
potential. But E391A should be viewed as the pilot study for the
more ambitious E14 proposal to the J-PARC facility, which aims at
eventually reaching below $10^{-12}$ sensitivity to probe BSM.
These are clean modes theoretically, so the challenge is for
experiment.

\section{\boldmath $\tau$: LFV and $(B-L)$V}
\label{sec:tau}

Before concluding, we touch upon exciting developments in rare tau
decays: radiative decays which have $b\to s$ echoes, and the
enigmatic (if found) baryon number violating decays.

\subsection{\boldmath $\tau\to \ell\gamma$, $\ell\ell\ell^\prime$}
\label{sec:tau_radiative}

The $\tau\to \ell\gamma$ processes are extremely suppressed in SM
by the very light neutrino mass. This opens up the opportunity to
probe BSM, just like the venerable $\mu\to e\gamma$ (where there
is the fabulous MEG experiment at PSI). Observation of lepton
flavor violating (LFV) decays would definitely mean New Physics!
Again, the favorite is SUSY, ranging from sneutrino-chargino
loops, exotic Higgs, $R$-parity violation, $\nu_R$ in SO(10), or
large extra dimensions (LED). Predictions for $\tau\to\mu\gamma$,
$\ell\ell\ell$, $\ell\ell\ell^\prime$, $\ell M^0$ (where $M^0$ is
a neutral meson) could reach the $10^{-7}$ level. The models are
often well motivated from observed near maximal
$\nu_\mu$-$\nu_\tau$ mixing, or interesting ideas such as
baryogenesis through leptogenesis. The great progress in neutrino
physics of the past decade has stimulated a lot of interest in
these LFV decays.

Experimentally, the stars are again the B factories: B factories
are also $\tau$ (and charm) factories, with $\sigma_{\tau\tau}
\sim 0.9\,$nb which is comparable to $\sigma_{bb} \sim 1.1\,$nb.
With steady increase in data, the B factories have pushed the
limits from $10^{-6}$ of the CLEO era, reaching down to
$10^{-8}$level. For example, with the 535 fb$^{-1}$ analysis by
Belle~\cite{tauto3l_Belle07} the limits on
$\tau\to\ell\ell\ell^\prime$ modes such as $\mu^+e^-e^-$ and
$e^+\mu^-\mu^-$ have reached $2\times 10^{-8}$, with BaBar not far
behind~\cite{tauto3l_BaBar07}. Thus, some models or the parameter
space are now ruled out.
%

To probe deeper into the parameter space of various LFV rare
$\tau$ decays, a Super B factory would be very helpful. In the
near future, LHCb can compete in the all charged track modes.

\subsection{\boldmath $\tau\to \Lambda\pi$, $p\pi^0$}
\label{sec:tau_BNV}

A somewhat wild idea is to search for baryon number violation
(BNV) in $\tau$ decay, i.e. involving the 3rd generation. This was
pointed out in Ref.~\cite{BNM05}, but the same reference argued
that, by linking to the extremely stringent limit on proton decay,
BNV ($B-L$ violating to be more precise) involving higher
generations are in general too small to be observed. This did not
stop Belle from conducting a search~\cite{BNM_Belle05}, followed
by BaBar~\cite{BNM_BaBar06}. So far, no signal is found, as
expected.

\section{Discussion and Conclusion}
\label{sec:Conclusion}

The last subsection brings us to``wilder" speculations, which we
have shunned so far. In the SUSY conference, however, ideas range
widely, if not wildly. To this author, from an experimental point
of view, the question is identifying the smoking gun, or else it
is better to stick to the simplest explanation of an effect that
requires New Physics. That has been our guiding principle.

Perhaps the wildest idea this year, and probably the one bringing
in the most insight, is about ``unparticle physics"~\cite{UnP07}.
Without discussing what this is, it has clearly stimulated much
interest. On the flavour and CPV front, for example, there is the
suggestion that unparticles could generate DCPV in unexpected
places~\cite{dcpvUnP07}. Sure enough, this observation may be
stimulated by the 3.2$\sigma$ indication~\cite{dcpvD+D-_Belle07}
of DCPV in $B^0\to D^-D^+$ by Belle (though the BaBar result is
consistent with zero~\cite{dcpvD+D-_BaBar07}) that is otherwise
very difficult to explain. But searching for DCPV in the $B^+\to
\tau^+\nu$ mode is also suggested~\cite{dcpvUnP07}, which is
interesting. If I may speculate, maybe unparticles could generate
BNV in the modes of the previous subsection. In any case, new
ideas such as these stimulate search efforts in otherwise
unmotivated places, hence are very valuable.

To summarize, I have covered a rather wide range of probes of TeV
scale physics via heavy flavour processes. At the moment, we have
two hints for New Physics: in the $\Delta {\cal S}$ difference
between TCPV in $B\to J/\psi K^0$ vs penguin dominant $b\to s\bar
qq$ modes; and in the experimentally established difference in
DCPV between $B^+\to K^+\pi^0$ and $B^0\to K^+\pi^-$ modes. These
are large CPV effects, but they are not unequivocal, either in
experimentation, or in interpretation. Because of this, the thing
to watch in 2008-2009, in my opinion, is whether Tevatron could
see a hint for {\it large} mixing-dependent CPV in $B_s\to
J/\psi\phi$, which would be unequivocal as evidence for BSM. If a
hint is seen, it can be quickly confirmed by LHCb. If Tevatron
fails to see any indication for $\sin2\Phi_{B_s}$, LHCb can probe
down to SM expectation rather quickly, but things would become
more and more boring. Other processes emphasized in this report
that has good potential for New Physics search are: direct CPV in
$B^+\to J/\psi K^+$; $B\to\tau\nu$; $b\to s\gamma$; ${\cal
A}_{FB}(B\to K^*\ell^+\ell^-$); $B_s\to \mu\mu$; $D^0$ mass mixing
and CPV; and $\tau \to \ell\gamma$.

The B factories have not yet exhausted their bag of surprises, but
a Super B factory is needed to better cover all the above subjects
(except $B_s\to \mu\mu$). Before that, we will attain some new
heights with LHCb.

\vskip0.2cm \noindent{\bf Acknowledgement} \ I thank Belle
spokespersons, in particular Masa Yamauchi, for nominating me, and
Hans K\"uhn for showing special appreciation of the presentation
of the talk.

%
%

\end{document}